\documentclass[10pt,a4paper]{article}
\usepackage{amsmath,latexsym,amssymb,amsfonts}
\usepackage[dvips]{graphicx}
\usepackage{bm}

\begin{document}

\title{\textbf{Oscillating instantons as\\ homogeneous tunneling channels}}
\author{\textsc{Bum-Hoon Lee}$^{a,b}$\footnote{bhl@sogang.ac.kr},\;\;
\textsc{Wonwoo Lee}$^{a}$\footnote{warrior@sogang.ac.kr},\;\; and
\textsc{Dong-han Yeom}$^{a,c}$\footnote{innocent.yeom@gmail.com}\\
\textit{$^{a}$\small{Center for Quantum Spacetime, Sogang University, Seoul 121-742, Republic of Korea}}\\
\textit{$^{b}$\small{Department of Physics, Sogang University, Seoul 121-742, Republic of Korea}}\\
\textit{$^{c}$\small{Yukawa Institute for Theoretical Physics, Kyoto University, Kyoto 606-8502, Japan}}}
\maketitle

\begin{flushright}
{\tt YITP-13-42}
\end{flushright}

\begin{abstract}
In this paper, we study Einstein gravity with a minimally coupled scalar field accompanied with a potential, assuming an $O(4)$ symmetric metric ansatz. We call an Euclidean instanton is to be an oscillating instanton, if there exists a point where the derivative of the scale factor and the scalar field vanish at the same time. Then we can prove that the oscillating instanton can be analytically continued, both as inhomogeneous and homogeneous tunneling channels.

Here, we especially focus on the possibility of a homogeneous tunneling channel. For the existence of such an instanton, we have to assume three things: (1) there should be a local maximum and the curvature of the maximum should be sufficiently large, (2) there should be a local minimum and (3) the other side of the potential should have a sufficiently deeper vacuum. Then, we can show that there exists a number of oscillating instanton solutions and their probabilities are higher compared to the Hawking-Moss instantons.

We also check the possibility when the oscillating instantons are comparable with the Coleman-DeLuccia channels. Thus, for a general vacuum decay problem, we should not ignore the oscillating instanton channels.\\

\end{abstract}

\newpage

\tableofcontents

\newpage

\section{Introduction}

In the context of inflation, the first order phase transition has been attributed to be an important physical aspect. Although there are many issues to be addressed, we will only focus on some of them. First, it was speculated that inflation will be finished by a first order phase transition \cite{Guth:1980zm,Sato:1980yn}; although it was noticed later that it is not sufficient to end inflation \cite{Guth:1982pn}. Second, even though it is not so useful for the primordial inflation, if there is secondary inflation for some cosmological purposes \cite{Lyth:1995ka}, then the first order phase transition can be quite important to end the process. This leads to some observational signatures \cite{Easther:2008sx}. Third, in the context of cosmic landscape and multiverse \cite{Susskind:2003kw}, the first order phase transition is indeed the mechanism to populate all vacua during the eternal inflation. Therefore, to understand the probability assignments of the multiverse (measure of the multiverse), we need to know about the various tunneling processes. Fourth, in string cosmology, there exist some phenomenological models with flux compactification \cite{Kachru:2003aw}. The question is whether such a model is viable for realistic cosmology or not. One of the criterion for that, at length, is the existence of a de Sitter vacuum with sufficient lifetime. To understand this problem, we have to know all the possible processes of vacuum decay. Fifth, if we live in a large bubble and if the bubble collision happens with a large size (larger than the Hubble size), then it can cause a kind of anisotropy in the cosmological signatures, and it may be observed \cite{Chang:2007eq,Chang:2007eq2}.

The vacuum transition process is a non-perturbative phenomena. The useful way to make sense of such non-perturbative problem is by taking recourse to Euclidean method. For general problems, the initial state can be approximated by the ground state. Then, the ground state wave function can be described by an \textit{Euclidean path intengral} \cite{Hartle:1983ai}, and this path integral can be approximated as a sum over all \textit{instantons} \cite{Hartle:2007gi,Hartle:2008ng}. Therefore, the usual way to find vacuum transition channels is as follows: (1) find an instanton and (2) analytically continue it to the Lorentzian signatures and (3) finally check the subsequent time evolutions.

In the literature, we are familiar with two famous Euclidean instantons for the vacuum decay process. For example, we know the \textit{Coleman-DeLuccia instantons} for inhomogeneous tunneling \cite{Coleman:1980aw} and the \textit{Hawking-Moss instantons} for homogeneous tunneling \cite{Hawking:1981fz}. The Coleman-DeLuccia instanton describes the tunneling between two local minima. Initially, all fields are in the false vacuum, and after the transition, a small true vacuum region appears. Between the two regions, there is a gradient of the field and it works as a wall with tension. If the wall can be made sufficiently thin, we can make use of the thin-wall approximation, and in this case, we can write an analytic formula for the probability. On the other hand, the Hawking-Moss instanton describes the tunneling from a false vacuum region to the local maximum of the potential. One interesting point to remember is that the tunneling seems to happen homogeneously. Of course, it does not imply that the whole universe (possibly of infinite size) tunnels to the local maximum at the same time; it does not make any sense. A more reasonable interpretation comes through the stochastic approach \cite{Linde:1993xx}. If short length modes of quantum fluctuations can be sufficiently suppressed, then the field dynamics smaller than the Hubble scale can be approximated with a homogeneous random walking of the field. This is described by the Langevin equation and, similar to the case of a the Brownian motion, all the possible ensembles will now give a statistical distribution. The statistical distribution will be described by the Fokker-Planck equation. One impressive result is that the stationary solution of the Fokker-Planck equation may correspond to the probability of the Hawking-Moss instantons. This explains why the Hawking-Moss instanton \textit{looks like} a homogeneous tunneling process.

We are aware of many examples with possible extensions of the Coleman-DeLuccia type instantons. We can consider not only the small bubbles (smaller than the background cosmological horizon) but also the large bubbles (larger than the background cosmological horizon) \cite{LW}. One can consider modified gravity and small false vacuum bubbles in the Jordan frame \cite{Lee:2006vka,Lee:2006vka2,Lee:2006vka3,Lee:2006vka4}. False vacuum bubbles are related to the bubble universe \cite{Hwang:2010gc,Hwang:2010gc2,Hwang:2010gc3}. One can also observe the tunneling of the tachyonic potential, where in general this cannot be described by the thin-wall approximation \cite{Lee:2012ug}. If gravitational effects become significant, then one can encounter non-trivial instantons due to gravitation.

However, if we compare this development with the homogeneous tunneling channel, we find very few applications that extend the Hawking-Moss type instantons. If we analytically continue all field values to be complexified functions, as in the case of the no-boundary measure \cite{Hartle:2007gi,Hartle:2008ng,Hwang:2011mp,Hwang:2012zj}, then we may obtain more instanton solutions. Before studying this process, in this paper, we will show that there are still quite a few real valued (not complex valued) instanton solutions with non-trivial field dynamics as homogeneous tunneling channels.

This type of solution has already been investigated in the literature \cite{Hackworth:2004xb,Brown:2007sd}. However, in those papers, they described the instanton solutions in an approximated way: they fixed a gravitational background, and considered the scalar field dynamics only. The dynamics with full gravitational back-reactions was studied by the authors \cite{Lee:2011ms}. However, in the previous literature, the authors only considered the solutions as inhomogeneous tunneling channels, while the present paper we will focus on the homogeneous channels.

In this paper, we will call our new instanton solutions as the \textit{oscillating instantons} (contrast to oscillating bounces \cite{Hackworth:2004xb,Brown:2007sd}, where these are not homogeneous in general). We use both analytic and numerical ways to describe the field dynamics and the gravitational back reactions at the same time. We will now discuss
\begin{itemize}
\item the definition and possible analytic continuations (Section~\ref{sec:def}),
\item conditions for existence and properties of oscillating instantons (Sections~\ref{sec:pro} and \ref{sec:gen}), and
\item finally, the comparison with other tunneling channels (Section~\ref{sec:app}).
\end{itemize}

\section{\label{sec:pre}Preliminaries}

In this section, we will discuss the basics of homogeneous tunneling channels. The most common one is the Hawking-Moss instanton. After complexification, we can generalize them to the fuzzy instantons. In this paper, we will discuss instantons that are not quite the same as the Hawking-Moss instantons or the fuzzy instantons, either. In this paper, we fix $c=G=\hbar =1$.

\subsection{Hawking-Moss instantons}

The wave function of the universe to describe its ground state is as follows \cite{Hartle:1983ai}:
\begin{eqnarray}\label{eqn:noboundary}
\Psi[h_{\mu\nu}, \chi] = \int_{\partial g = h, \partial \phi = \chi} \mathcal{D}g\mathcal{D}\phi \;\; e^{-S_{\mathrm{E}}[g,\phi]},
\end{eqnarray}
where $h_{\mu\nu}$ and $\chi$ are the boundary values of the Euclidean metric $g_{\mu\nu}$ and the matter field $\phi$ which are the integration variables and the integration is over all non-singular geometries with a single boundary. Here, we will consider the Euclidean action
\begin{eqnarray}\label{eq:action}
S_{\mathrm{E}} = -\int d^{4}x \sqrt{+g} \left( \frac{1}{16\pi}R - \frac{1}{2} (\nabla \phi)^{2} - V(\phi) \right).
\end{eqnarray}

We choose the minisuperspace approximation:
\begin{eqnarray}\label{eq:mini}
ds_{\mathrm{E}}^{2} = d\eta^{2} + \rho^{2}\left(\eta\right) \left(d\chi^{2} + \sin^{2}\chi \left( d\theta^{2} + \sin^{2}\theta d\varphi^{2} \right) \right),
\end{eqnarray}
and then the Euclidean action reduces to
\begin{eqnarray}
S_{\mathrm{E}} = 2 \pi^{2} \int d\eta \left[ -\frac{3}{8\pi} \left( \rho \dot{\rho}^{2} + \rho \right) + \frac{1}{2}\rho^{3} \dot{\phi}^{2} + \rho^{3} V(\phi) \right].
\end{eqnarray}

We use the steepest-descent method and consider only the on-shell solutions to count the probability from the path-integral \cite{Hartle:2007gi,Hartle:2008ng,Hwang:2011mp,Hwang:2012zj}. We solve the classical equations of motion for Euclidean as well as Lorentzian time directions
\begin{eqnarray}
\label{E3}\ddot{\phi} &=& - 3 \frac{\dot{\rho}}{\rho} \dot{\phi} \pm V',\\
\label{E4}\ddot{\rho} &=& - \frac{8 \pi}{3} \rho \left( \dot{\phi}^{2} \pm V \right),
\end{eqnarray}
where the upper sign is for the Euclidean time and the lower one is for the Lorentzian time.
Then the on shell Euclidean action is given by
\begin{eqnarray}
S_{\,\mathrm{E}} = 4\pi^{2} \int d \eta \left( \rho^{3} V - \frac{3}{8 \pi} \rho \right).
\end{eqnarray}

When $V'(\phi_{0})=0$, we can find an analytic solution:
\begin{eqnarray}
\phi &=& \phi_{0}\\
\rho &=& \frac{1}{H} \sin H \eta,
\end{eqnarray}
where $H=\sqrt{8\pi V(\phi_{0})/3}$. Here, we have used the regular initial conditions
\begin{eqnarray}\label{eq:initial}
\rho(0)=0, \;\;\; \dot{\rho}(0)=1, \;\;\; \dot{\phi}(0)=0,
\end{eqnarray}
at $\eta=0$.

We want to analytically continue the solution to the Lorentzian manifold using $d\eta = i dt$. Then at the turning point $\eta = \eta_{\mathrm{max}}$, we have to impose the following
\begin{align}\label{eq:turn}
\rho(t=0) = \rho(\eta=\eta_{\mathrm{max}}), \;\;\;\; \dot{\rho}(t=0)=i\dot{\rho}(\eta=\eta_{\mathrm{max}}),\\
\label{eq:turn2}
\phi(t=0) = \phi(\eta=\eta_{\mathrm{max}}), \;\;\;\; \dot{\phi}(t=0)=i\dot{\phi}(\eta=\eta_{\mathrm{max}}),
\end{align}
as derived from the Cauchy-Riemann condition. Unless $\dot{\rho}(\eta_{\mathrm{max}}) = \dot{\phi}(\eta_{\mathrm{max}}) = 0$, we should have complex valued functions for $\phi$ and $\rho$ for the Lorentzian time $t$. Therefore, we can analytically continue the solution only at $\eta_{\mathrm{max}} = \pi/2H$; thus preserving the real valuedness of the functions for all $\eta$ and $t$.

In this particular case, the on-shell action becomes
\begin{eqnarray}
S_{\mathrm{E}} = - \frac{3}{16 V(\phi_{0})}.
\end{eqnarray}
If this instanton mediates a tunneling from a local minimum $\phi_{\mathrm{m}}$ to a local maximum $\phi_{\mathrm{M}}$ of a potential, then we obtain the tunneling probability to be,
\begin{eqnarray}
P \cong \exp \left( \frac{3}{8 V(\phi_{\mathrm{M}})} -  \frac{3}{8 V(\phi_{\mathrm{m}})} \right).
\end{eqnarray}
This is known as the Hawking-Moss instanton \cite{Hawking:1981fz}.

\subsection{Generalization to fuzzy instantons}

We can generalize our solution to complex valued functions. Of course, for a long enough Lorentzian time, we expect all functions to be real. This condition is related to the classicality condition \cite{Hartle:2007gi,Hartle:2008ng}. Due to the analytic continuation to complex functions, the action is in general complex, so that
\begin{equation}
\Psi[a,\chi] = A[a,\chi] e^{i S[a,\chi]}
\label{eq:class}
\end{equation}
with $A,S$ being real. If the rate of change of $S$ is much greater than that of $A$, i.e.,
\begin{equation} \label{eqn:classicality}
|\nabla_I A(q)|\ll |\nabla_I S(q)|, \qquad I=1,\ldots n,
\end{equation}
where $I$ denotes the index of the canonical variables, then the wave function describes an almost classical behavior \cite{Hartle:2007gi,Hartle:2008ng,Hwang:2011mp,Hwang:2012zj}.

Let us consider the situation when all functions $\rho$ and $\phi$ are complexified by $\rho^{\mathfrak{Re}}+i\rho^{\mathfrak{Im}}$ and $\phi^{\mathfrak{Re}}+i\phi^{\mathfrak{Im}}$. Then the required initial conditions for regularity are
\begin{eqnarray}
\rho(0)^{\mathfrak{Re}} = \rho(0)^{\mathfrak{Im}} &=& 0,\\
\dot{\rho}(0)^{\mathfrak{Re}} &=& 1,\\
\dot{\rho}(0)^{\mathfrak{Im}} &=& 0,\\
\dot{\phi}(0)^{\mathfrak{Re}} = \dot{\phi}(0)^{\mathfrak{Im}} &=& 0.
\end{eqnarray}
At the junction time $\eta = \eta_{\mathrm{max}}$, we paste $\rho(\eta)$ and $\phi(\eta)$ to $\rho(t)$ and $\phi(t)$ using Equations~(\ref{eq:turn}) and (\ref{eq:turn2}). The remaining initial conditions are the initial field value $\phi(0) = \phi_{0} e^{i\theta}$, where $\phi_{0}$ is positive valued and $\theta$ is a phase angle. After fixing $\phi_{0}$, by tuning the two parameters $\theta$ and the turning point $\eta_{\mathrm{max}}$, one can find a classcialized instanton solution \cite{Hwang:2011mp}\cite{Hwang:2012zj}. If there exists a classical history, then we may calculate a meaningful probability for the classical universe.

Around the local maximum of a tachyonic potential
\begin{equation}
V(\phi) = V_{0} - \frac{1}{2} m^{2} \phi^{2},
\end{equation}
as appeared in the authors' previous paper \cite{Hwang:2012zj}, we numerically classified all fuzzy instantons around the local maximum, by defining $\mu^{2} = m^{2}/V_{0}$.
\begin{itemize}
\item Slow-rolling case: $\mu^{2} \lesssim 1$
\begin{eqnarray}
P \cong \int \exp{\left[\frac{3}{8}\left(\frac{1}{V(\phi_{0})}-\frac{1}{V_\mathrm{m}}\right) \right]} d\phi_{0}
\end{eqnarray}
\item Fast-rolling case: $\mu^{2} \gtrsim 10$
\begin{eqnarray}
P \cong \exp{\left[\frac{3}{8}\left(\frac{1}{V_{0}}-\frac{1}{V_\mathrm{m}}\right) \right]} \Delta \phi_{0}
\end{eqnarray}
\end{itemize}
Here, the allowed field space is
\begin{equation}
\Delta \phi_0 \propto \begin{cases}
\mu^{-1} & \text{for\;\;\;} \mu^{2} \lesssim 1\\
\exp{- c \mu}& \text{for\;\;\;} \mu^{2} \gtrsim 10
\end{cases}
\label{eq:deltaphi}
\end{equation}
with $c \simeq 2.11$. Therefore, we can conclude that if $\mu \gg 1$ around the local maximum, the field space which allows fuzzy instantons shrinks exponentially.

\subsection{Motivation for this paper}

\begin{description}
\item[A new channel of homogeneous tunneling:] Let us consider, for example, a double-well potential of the form
\begin{equation}\label{eq:pot}
V(\phi) = V_{0} \left( 1 - \frac{1}{2} \mu^{2} \phi^{2} + \frac{1}{4!} \lambda \phi^{4} \right),
\end{equation}
and consider a homogeneous tunneling from the left vacuum to the left/right vacuum.\footnote{In this paper, we consider the case where we were initially in the left vacuum and finally the field tunnels to the left or right vacuum; `from left vacuum to left vacuum' describes the situation with the field tunneling to a point near the local maximum, but it cannot go over the potential wall, and rolls back to the original vacuum.} What is commonly known is the Hawking-Moss instanton and recently the fuzzy instantons have been found.

In the present work, we will discuss quite a few solutions that mediates tunneling from the left vacuum to the left/right vacuum, where they have got non-trivial field dynamics (so, it is not the original Hawking-Moss instanton) and they are real valued functions (so, it is not the fuzzy instanton that we found). In general, such discrete solutions will be subdominant compared to the fuzzy instantons, since they have measure zero over the on-shell solution space, while the fuzzy instantons have non-zero measure. However, when $\mu^{2}$ becomes sufficiently large, $\Delta \phi_{0}$ will exponentially decrease. In this limit, if we compare using a proper measure including the off-shell space, then a discrete number of solutions are comparable with the fuzzy instantons with exponentially small field space. In this limit, our new solutions are thus quite useful as a non-trivial alternative to the vacuum transition.

\item[Academic interests:] From the literature, we know about various kinds of Euclidean $O(4)$-symmetric instantons or bounce solutions. However, till now, for a given instanton solution, we could apply only one of the analytic continuations to the Lorentzian manifold: the Coleman-DeLuccia type $\chi \rightarrow \pi/2 + it$ or the Hawking-Moss type $\eta \rightarrow \eta_{\mathrm{max}} + it$.

    In the present paper, we find an interesting example where the instanton can be interpreted in both ways: Coleman-DeLuccia type and Hawking-Moss type at the same time. The former case has already been applied and studied in depth as a bounce solution \cite{Lee:2011ms}. In this paper, we will re-interpret the same instanton using the Hawking-Moss type analytic continuation. Therefore, this work is of some academic interest.
\end{description}

\begin{figure}
\begin{center}
\includegraphics[scale=0.4]{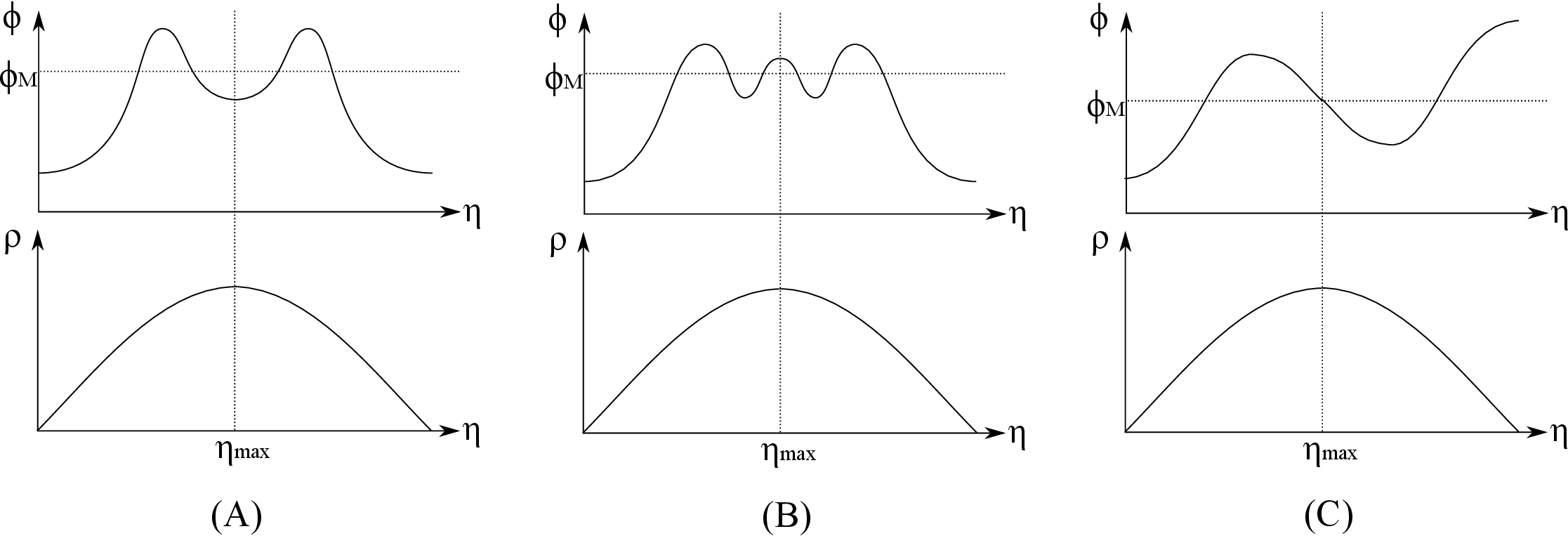}
\caption{\label{fig:field}(A) and (B) are examples of oscillating instanton solutions with oscillating number (number of local maxima of the scalar field before reaching $\eta_{\mathrm{max}}$) $1$ and $2$. It is also possible to think a solution like (C): in this case, the analytic continuation $\chi \rightarrow \pi/2 + it$ is possible but $\eta \rightarrow \eta_{\mathrm{max}} + i t$ is not possible.}
\end{center}
\end{figure}
\begin{figure}
\begin{center}
\includegraphics[scale=0.23]{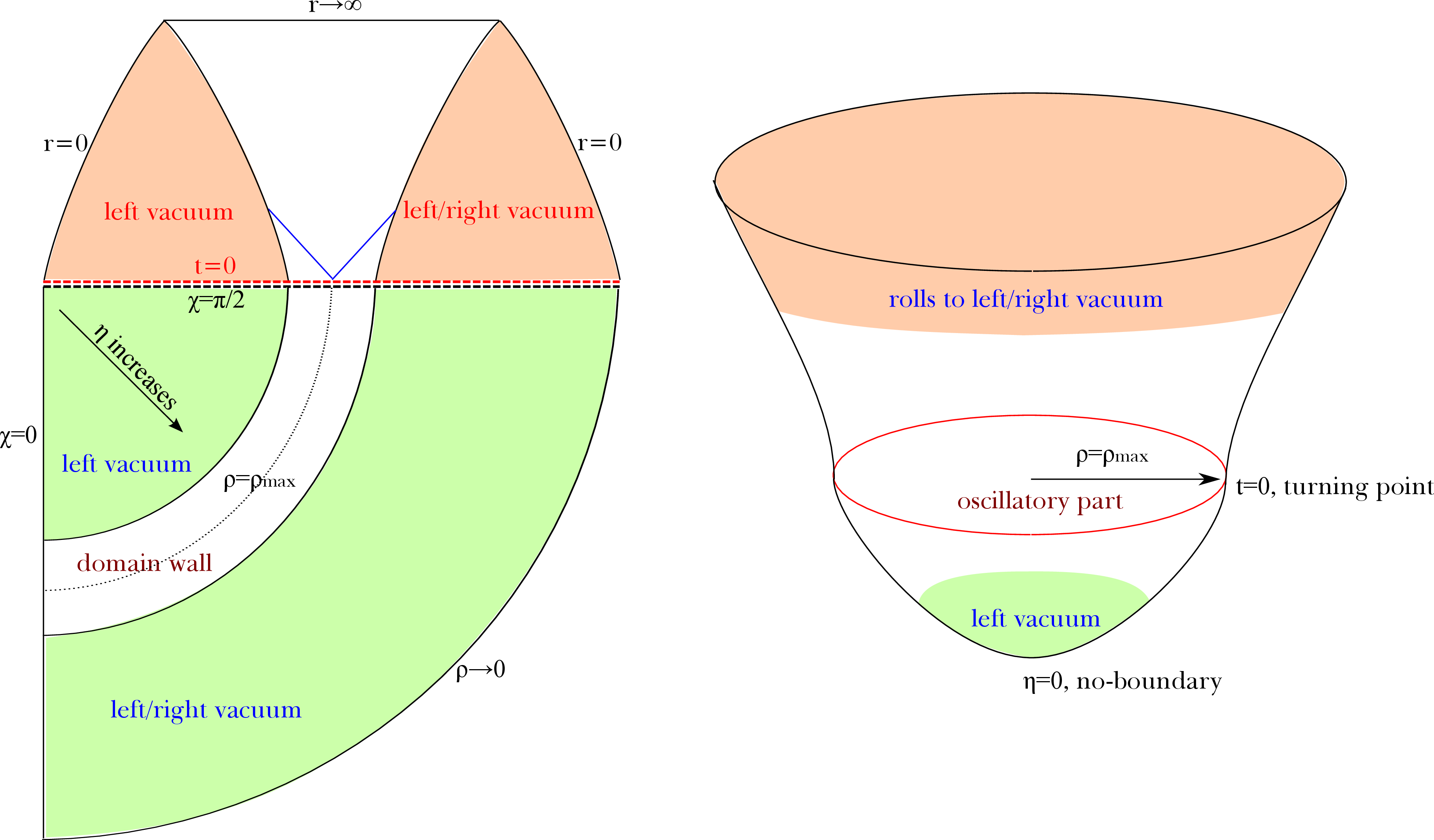}
\caption{\label{fig:interp1}Left: analytic continuation $\chi \rightarrow \pi/2 + it$. Right: analytic continuation $\eta \rightarrow \eta_{\mathrm{max}} + i t$.}
\end{center}
\end{figure}

\section{\label{sec:osc}Oscillating instanton solutions}

\subsection{\label{sec:def}Definition of oscillating instantons}

We will consider Euclidean Einstein gravity with a minimally coupled scalar field given by the action in Equation~(\ref{eq:action}). In addition, we assume the $O(4)$-symmetric metric ansatz as in Equation~(\ref{eq:mini}). Then the system of scalar field $\phi$ and metric $\rho$ is governed by two second order differential equations, namely Equations~(\ref{E3}) and (\ref{E4}), with initial regular conditions at $\eta = 0$ given by Equation~(\ref{eq:initial}).

\paragraph{Definition} \textit{A set of solutions for the Einstein and scalar field equations in Euclidean signature $\rho(\eta)$ and $\phi(\eta)$ is an \textbf{oscillating instanton}, if there exists an Euclidean time $\eta_{\mathrm{max}}>0$ such that $\dot{\rho}(\eta_{\mathrm{max}})=\dot{\phi}(\eta_{\mathrm{max}})=0$.\\}

We can observe the following two properties:
\begin{enumerate}
\item
Let us consider the field ansatz for $\rho(\eta)$ and $\phi(\eta)$ as follows:
\begin{eqnarray}
\rho(\eta) = \bar{\rho}(\eta), \;\;\; \phi(\eta) = \bar{\phi}(\eta), \;\;\; &\mathrm{for\;\;}& 0<\eta<\eta_{\mathrm{max}}\\
\rho(\eta) = \bar{\rho}(2 \eta_{\mathrm{max}} - \eta), \;\;\; \phi(\eta) = \bar{\phi}(2 \eta_{\mathrm{max}} - \eta), \;\;\; &\mathrm{for\;\;}& \eta_{\mathrm{max}}<\eta<2\eta_{\mathrm{max}},
\end{eqnarray}
where $\bar{\rho}(\eta)$ and $\bar{\phi}(\eta)$ is a solution for $0<\eta<\eta_{\mathrm{max}}$. Then, $\rho(\eta)$ and $\phi(\eta)$ are indeed solutions of the Einstein and field equations, if they satisfy $\dot{\rho}(\eta_{\mathrm{max}})=\dot{\phi}(\eta_{\mathrm{max}})=0$, because at the junction $\eta=\eta_{\mathrm{max}}$, the field values and their first derivatives should be continuous (e.g., (A) and (B) in Figure~\ref{fig:field}).

Therefore, for a given oscillating instanton solution for $0<\eta<\eta_{\mathrm{max}}$, we can extend it for $\eta_{\mathrm{max}}<\eta<2\eta_{\mathrm{max}}$. Now, by definition, $\rho(2\eta_{\mathrm{max}})=0$. In other words, if we analytically continue $\chi \rightarrow \pi/2 + it$, then the manifold has zero area regions ($r=0$) at $\eta = 0$ and $\eta = 2\eta_{\mathrm{max}}$. Therefore, we will infer that one domain is nucleated between two universes in the Lorentzian signature (Left in Figure~\ref{fig:interp1}).

\item If we turn to the Lorentzian time by $\eta \rightarrow \eta_{\mathrm{max}} + i t$, to satisfy the analyticity and reality conditions, we have to assume $\dot{\rho}(\eta_{\mathrm{max}})=\dot{\phi}(\eta_{\mathrm{max}})=0$. Therefore, for a given oscillating instanton solution, one can analytically continue it to the Lorentzian time by $\eta \rightarrow \eta_{\mathrm{max}} + i t$ at $\eta_{\mathrm{max}}$ (Right in Figure~\ref{fig:interp1}).
\end{enumerate}

There are few more comments to follow. (A) and (B) are examples of oscillating instantons with oscillating number (number of local maxima of the scalar field before reaching $\eta_{\mathrm{max}}$) $1$ and $2$.\footnote{This particular convention to label a number is different from that of \cite{Hackworth:2004xb}; they counted the number by counting the number of fields crossing the local maximum of the potential.} For odd number of oscillation, after the $\eta \rightarrow \eta_{\mathrm{max}} + i t$ rotation, the field rolls to the original vacuum (the field at the turning point is the one at the left side of $\phi_{\mathrm{M}}$); for even number of oscillation, after the $\eta \rightarrow \eta_{\mathrm{max}} + i t$ rotation, the field rolls to the other side of the vacuum (the field at the turning point is the one at the right side of $\phi_{\mathrm{M}}$). For the cases (A) and (B) depicted in Figure~\ref{fig:field}, to justify both the analytic continuations of Left and Right of Figure~\ref{fig:interp1}, what we need is the definition of the oscillating instantons (in other words, it does not depend on the conditions of the potential further). One interesting extension of the case is (C) in Figure~\ref{fig:field}: here, we need a certain symmetry for the potential (e.g., $V(\phi-\phi_{\mathrm{M}})=V(\phi+\phi_{\mathrm{M}})$), and the analytic continuation $\chi \rightarrow \pi/2 + it$ is possible but $\eta \rightarrow \eta_{\mathrm{max}} + i t$ is not possible. In this paper, we will not discuss the details of type (C).

\subsection{\label{sec:pro}Properties of oscillating instantons: Double-well potential}

To study the existence and the properties of oscillating instantons, we first study the simplest case given by Equation~(\ref{eq:pot}). Here, one can define the following rescaling of the variables,
\begin{eqnarray}
d\eta &\rightarrow& \frac{d\eta}{\sqrt{V_{0}}},\\
\rho &\rightarrow& \frac{\rho}{\sqrt{V_{0}}}.
\end{eqnarray}
Then we can obtain the rescaled action to be
\begin{eqnarray}
S_{\mathrm{E}} \rightarrow \frac{S_{\mathrm{E}}}{V_{0}}
\end{eqnarray}
and the rescaled potential to be
\begin{eqnarray}
V(\phi) \rightarrow \frac{V(\phi)}{V_{0}}.
\end{eqnarray}
Therefore, without any loss of generality, we can choose $V_{0} = 1$ and finally one can restore this back to the results. In this subsection, we will choose $V_{0}=1$.
The local minimum is at $\pm \sqrt{6\mu^{2}/\lambda}$. We choose the vacuum energy at the local minimum to be zero: $\lambda=3\mu^{4}/2$. We choose such that the potential becomes sufficiently steep around the local maximum $\mu^{2} \gg 1$.

\subsubsection{Around the local maximum: Maximum number of oscillations}

Around the local maximum, we can write the potential as
\begin{equation}
V(\phi) = 1 - \frac{1}{2} \mu^{2} \phi^{2}.
\end{equation}
When we consider fast-rolling fields around the local maximum, we may approximate the scalar field by $\phi \sim A \sin \omega \eta$. In this limit, we can reasonably approximate the Einstein equation by
\begin{eqnarray}
\dot{\rho}^{2} = 1 + \frac{8 \pi}{3} \rho^{2}.
\end{eqnarray}
If we define $H_{0}^{2} = 8\pi /3$, then the solution for the metric factor becomes
\begin{eqnarray}
\rho(\eta) = \frac{1}{H_{0}} \sin H_{0} \eta
\end{eqnarray}
which satisfies the initial conditions $\rho(0)=0$ and $\dot{\rho}=1$ for regularity. Then $\dot{\rho}(\pi/2H_{0})= 0$ is a plausible candidate of the turning time to paste the Lorentzian manifold $dt = -i d\eta$.

Now we can write the equation for the fields as follows:
\begin{eqnarray}
\ddot{\phi} + 3 H_{0} \dot{\phi} \cot H_{0}\eta + \mu^{2} \phi = 0.
\end{eqnarray}
Note that
\begin{eqnarray}
\cot H_{0}\eta \rightarrow 0
\end{eqnarray}
around the turning point $\eta = \pi/2H_{0}$. Therefore for $H_{0} t \rightarrow \pi/2$,
\begin{eqnarray}
\phi(\eta) = C_{1} \cos \left(\mu\eta + \delta \right),
\end{eqnarray}
where $C_{1}$ and $\delta$ are determined by initial conditions.
The oscillating frequency around the turning point is $\mu$; hence, the maximum number of oscillation $N$ before the turning time will be approximately given by
\begin{eqnarray}
N \lesssim \frac{2 \times (\mathrm{turning\; time})}{(\mathrm{oscillation\; period})} \thickapprox \frac{\mu}{2H_{0}} \thickapprox 0.17 \mu.
\end{eqnarray}
Therefore, for the existence of oscillating instantons, we require $\mu > 5.9$ or $\mu^{2} > 35$ \cite{Hackworth:2004xb}.

\begin{figure}
\begin{center}
\includegraphics[scale=0.5]{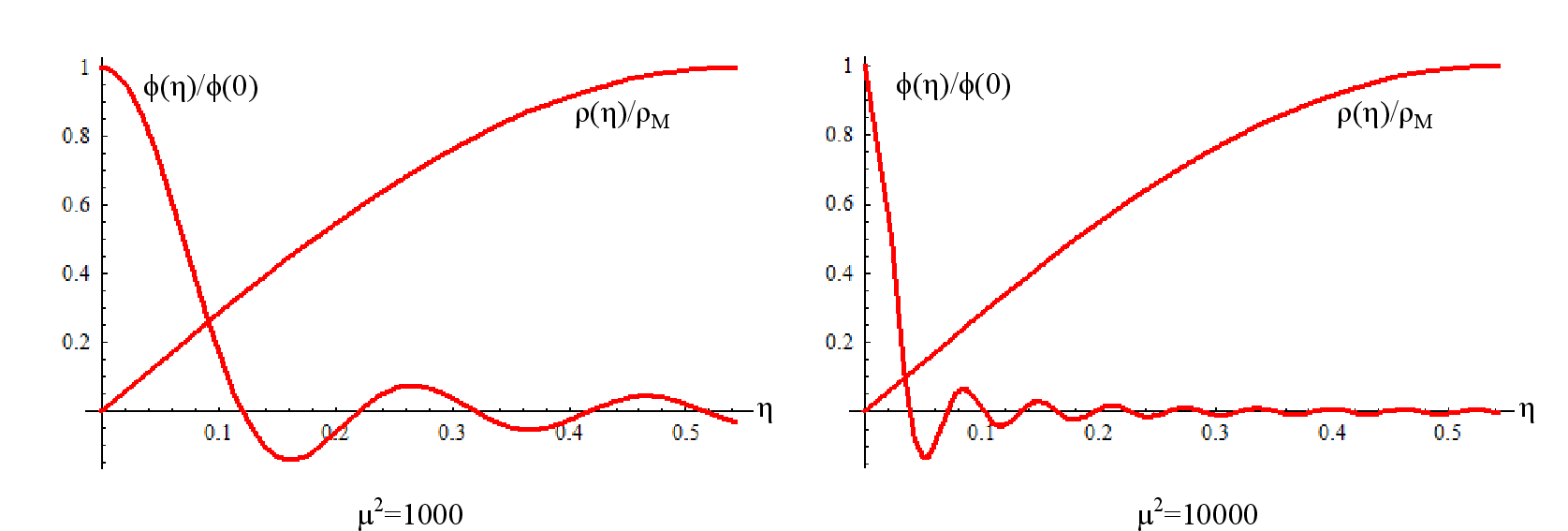}
\caption{\label{fig:maxosc}Solutions for $\phi(0)=0.0001 \times \sqrt{6\mu^{2}/\lambda}$ suchthat the initial condition is near the local maximum. Two plots denote $\phi(\eta)/\phi(0)$ and $\rho(\eta)/\rho_{\mathrm{M}}$, where $\rho_{\mathrm{M}}=\sqrt{3/8\pi}$. $\dot{\rho}=0$ around $\eta \sim 0.54$. If there are $\dot{\phi}=0$ points for $\eta < 0.54$, then there is a hope to see oscillating instantons by tuning $\phi(0)$. For $\mu^{2}=1000$, at least $4$ peaks appear for $\phi$; for $\mu^{2}=10000$, at least $16$ peaks appear for $\phi$.}
\end{center}
\end{figure}

Figure~\ref{fig:maxosc} denotes solutions for $\phi(0)=0.0001 \times \sqrt{6\mu^{2}/\lambda}$ such that the initial condition is near the local maximum. Two plots denote $\phi(\eta)/\phi(0)$ and $\rho(\eta)/\rho_{\mathrm{M}}$, where $\rho_{\mathrm{M}}\equiv\sqrt{3/8\pi}$. $\dot{\rho}=0$ appears around $\eta \sim 0.54$. If there are $\dot{\phi}=0$ points for $\eta < 0.54$, then there are still some hope to see the oscillating instanton solutions by tuning $\phi(0)$. For $\mu^{2}=1000$, at least $4$ peaks appear for $\phi$; for $\mu^{2}=10000$, at least $16$ peaks appear for $\phi$. Thus, as we tune the initial condition $\phi(0)$, there are some hope to see $4$ or $16$ oscillating instantons in principle. Note that our previous analytic estimations for maximum number of oscillations are pretty good: for $\mu^{2}=1000$, the maximum number is $0.17 \times \mu \sim 5.3$; for $\mu^{2}=10000$, the maximum number is $0.17 \times \mu \sim 17$.

\subsubsection{Around the local minimum: Existence of oscillating instantons}

Around the local minimum, we can approximate the potential by
\begin{equation}
V(\phi) = \frac{1}{2} M^{2} \left(\phi-\phi_{\mathrm{m}}\right)^{2}
\end{equation}
with a certain mass scale $M$ around the local minimum. We assume
\begin{eqnarray}
\rho(\eta) \simeq \eta
\end{eqnarray}
and hence it is approximately flat.

Then the scalar field equation becomes
\begin{eqnarray}
\ddot{\phi} + \frac{3}{\eta} \dot{\phi} - M^{2} \left( \phi-\phi_{\mathrm{m}} \right) = 0.
\end{eqnarray}
Thus the general solution is given by
\begin{eqnarray}
\phi(\eta) = \phi_{\mathrm{m}} + C_{2} \frac{J_{1} (-i M \eta)}{\eta} + C_{3} \frac{Y_{1} (-i M \eta)}{\eta},
\end{eqnarray}
where $C_{2}$ and $C_{3}$ are determined by initial conditions. For the finiteness at $\eta=0$, we have to choose $C_{3}=0$ and the solution for $\eta \ll 1$ can be approximated as
\begin{eqnarray}
\frac{\phi(\eta)}{\phi(0)} = 1 - \frac{M^{2}}{8} \left( \frac{\phi_{\mathrm{m}} - \phi(0)}{\phi(0)} \right) \eta^{2} + \mathcal{O}\left( \eta^{4} \right).
\end{eqnarray}
Therefore, as $\phi(0)$ approaches $\phi_{\mathrm{m}}$, the scalar field slowly rolls up. By tuning the initial condition $\phi(0)$, we can in principle control the scalar field to have arbitrarily long time around the local minimum.

\begin{figure}
\begin{center}
\includegraphics[scale=0.7]{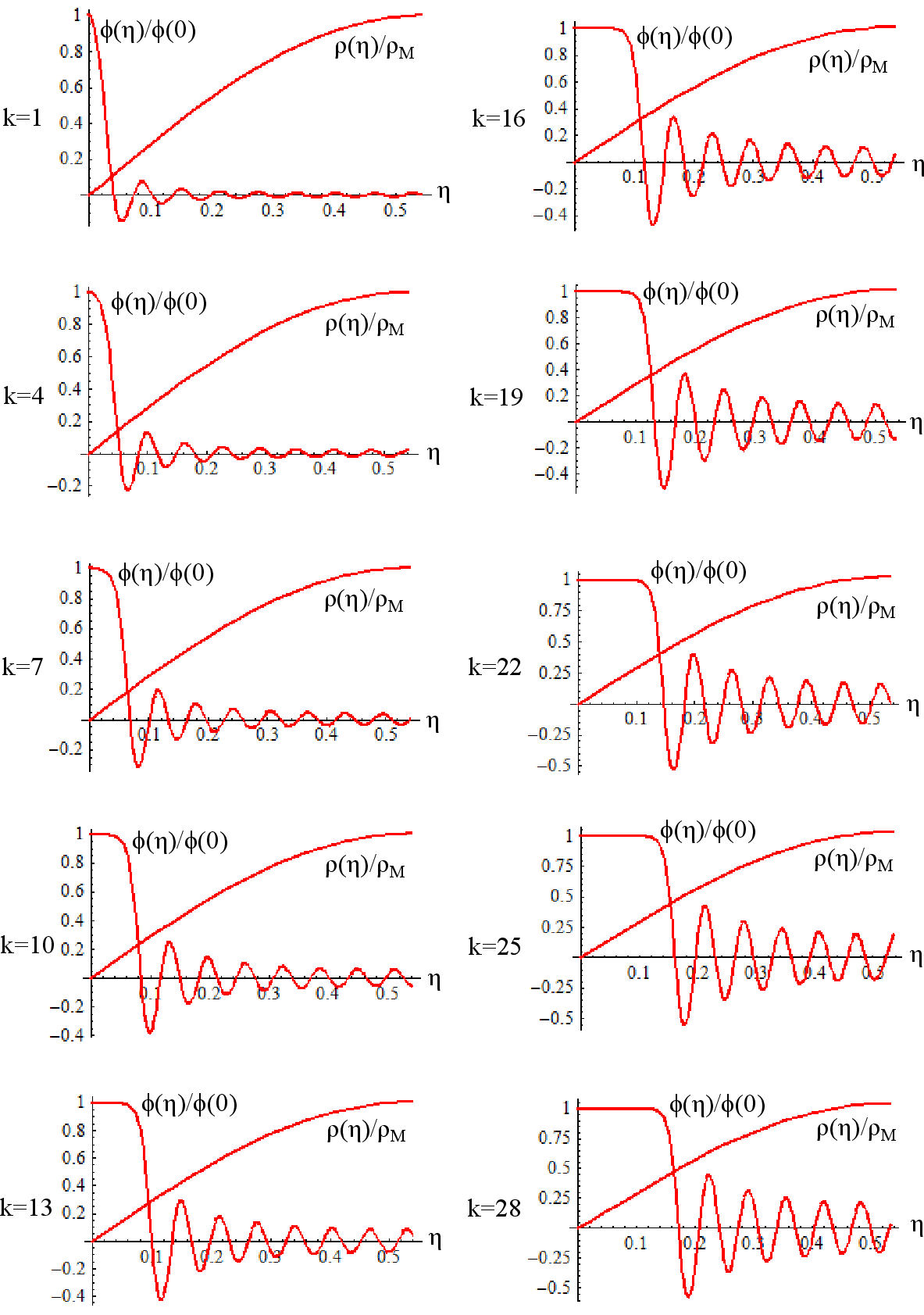}
\caption{\label{fig:mu2=10000}Solutions for $\mu^{2}=10000$. By varying $k$, where $\phi(0)=\sqrt{6\mu^{2}/\lambda} (1-2^{-k})$. As $k$ increases, the initial condition exponentially approaches the local minimum $\sqrt{6\mu^{2}/\lambda}$.}
\end{center}
\end{figure}

In Figure~\ref{fig:mu2=10000}, for a given $\mu^{2}=10000$, we vary initial conditions $\phi(0)=\phi_{\mathrm{m}} (1-2^{-k})$, where $\phi_{\mathrm{m}}=\sqrt{6\mu^{2}/\lambda}$. As $k$ increases, the initial condition exponentially approaches the local minimum. For small $k$, as was the case in Figure~\ref{fig:maxosc}, we can see $16$ extrema for $\phi$. For small $\eta$, the initial condition approaches the neighborhood around the local minimum and the scalar field spends longer and longer time near the local minimum. For sufficiently large $k$, for example $k=28$, we can see that the number of local extrema is $11$.

\begin{figure}
\begin{center}
\includegraphics[scale=1]{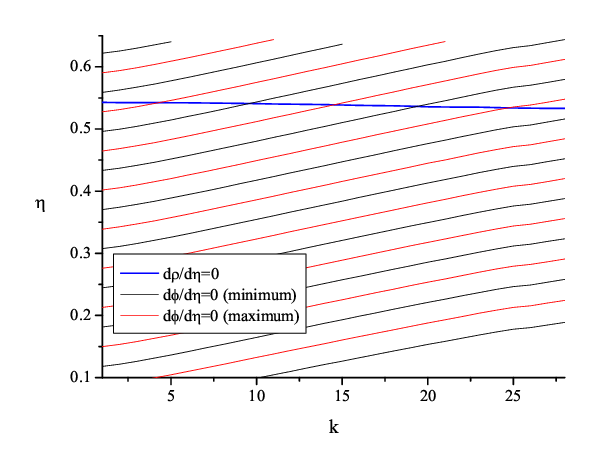}
\caption{\label{fig:solution}The location $\eta$ for $\dot{\rho}=0$ (blue) and $\dot{\phi}=0$ (red, black) by varying $k$, where $\phi(0)=\sqrt{6\mu^{2}/\lambda} (1-2^{-k})$. When the blue and red/black curves do coincide, there exists an oscillating instanton. Black curves stand for local minima while red curves stand for local maxima of $\phi(\eta)$.}
\end{center}
\end{figure}

Therefore, between $k=1$ and $k=28$, the local maximum of $\rho$ and the local extrema of $\phi$ coincide $16-11=5$ times. This is explicitly represented by Figure~\ref{fig:solution}. The blue curve is the location ($\eta$) of the local maximum of $\rho$ (i.e., $\eta$ for $\dot{\rho}(\eta)=0$) as a function of $k$. The red and black curves are the locations of the local extrema of $\phi$ (i.e., $\eta$ for $\dot{\phi}(\eta)=0$) as a function of $k$. Here, black curves denote the local minima and red curves stand for local maxima. Therefore, if the blue curve and the black/red curve coincide, the corresponding initial condition gives an oscillating instanton solution.

Both the Figures~\ref{fig:mu2=10000} and \ref{fig:solution} clearly show that more the initial condition approaches the local minimum, more does the field spend the time around the local minimum. Then, the blue curve and the first local extrema for $\phi$ will eventually meet for sufficiently large $k$. Therefore, this can confirm that \textit{if a double-well potential allows for maximum number of oscillations $N$, then there exist $N$ number of oscillating instantons by tuning the initial condition.}

\subsubsection{Hierarchy of probabilities}

Let us denote the Euclidean time of the first peak of the scalar field by $\eta_{0}$. We can approximate this solution as follows. (Here, we want to calculate the Euclidean action approximately; therefore, we do not have to worry about the continuity of the solution.):
\begin{itemize}
\item $\eta \lesssim \eta_{0}$
\begin{eqnarray}
\phi(\eta) &\simeq& \sqrt{6\mu^{2}/\lambda},\\
\rho(\eta) &\simeq& \eta.
\end{eqnarray}
\item $\eta \gtrsim \eta_{0}$
\begin{eqnarray}
\phi(\eta) &\simeq& 0,\\
\rho(\eta) &\simeq& \frac{1}{H_{0}} \sin H_{0} \eta,
\end{eqnarray}
where $H_{0}=\sqrt{8\pi V_{0}/3}$.
\end{itemize}
Also, one can observe from Figure~\ref{fig:solution} that $\eta_{0} \propto k$ where $\phi(0)=\sqrt{6\mu^{2}/\lambda} (1-2^{-k})$.

Therefore, the Euclidean action becomes approximately ($\eta_{0} H_{0} \lesssim 1$)
\begin{eqnarray}
S_{\mathrm{E}} \simeq 4\pi^{2} \int_{0}^{\eta_{0}} \left( -\frac{3}{8\pi} \eta \right) d\eta
+ 4\pi^{2} \int_{\eta_{0}}^{\pi/2H_{0}} \left( \rho^{3} V_{0} - \frac{3}{8\pi} \rho \right) d\eta.
\end{eqnarray}
For convenience, we define the action difference by $\Delta S_{\mathrm{E}}$ such that
\begin{eqnarray}
S_{\mathrm{E}} -\left(- \frac{3}{16V_{0}} \right) \equiv \Delta S_{\mathrm{E}} &\simeq& - \frac{3\pi}{4} \eta_{0}^{2} - \frac{3}{16V_{0}} \left( \cos^{3} H_{0}\eta_{0} - 1 \right)\\
&\simeq& - \frac{7}{6} \pi^{2} V_{0} \eta_{0}^{4} + \mathcal{O}\left(\left(H_{0}\eta_{0}\right)^{6}\right).
\end{eqnarray}

\begin{figure}
\begin{center}
\includegraphics[scale=1]{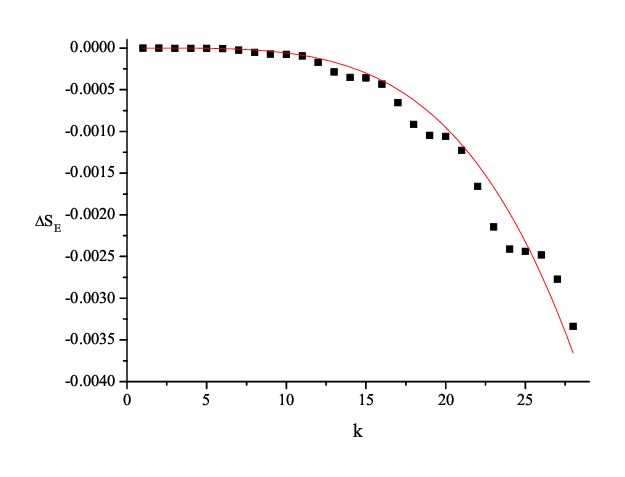}
\caption{\label{fig:action}The Euclidean action difference $\Delta S_{\mathrm{E}}$ as a function of $k$. The red curve fits the function $\Delta S_{\mathrm{E}} = C k^{4}$ with $C \simeq (-5.9 \pm 0.17) \times 10^{-9}$.}
\end{center}
\end{figure}

When we compare this with the numerical calculations ($\mu^{2}=10000$), the relation between $\eta_{0}$ and $k$ is approximately linear: $\eta_{0} \simeq 0.0052 k$. Therefore, we expect that the action difference will be approximately given by $\Delta S_{\mathrm{E}} \sim - 8.42 \times 10^{-9} \times k^{4}$. Figure~\ref{fig:action} contains the numerical results of the action difference $\Delta S_{\mathrm{E}}$ and we can notice that our rough estimation is not bad: $\Delta S_{\mathrm{E}} \simeq C k^{4}$, where $C \simeq (-5.9 \pm 0.17) \times 10^{-9}$.

For larger $\eta_{0}$ limit (or larger $k$ limit), the action will approximately increase to
\begin{eqnarray}
S_{\mathrm{E}} &\simeq& - \frac{3\pi}{4}\left( \frac{\pi}{2H_{0}} \right)^{2} = - \frac{9 \pi^{2}}{128 V_{0}},\\
\frac{S_{\mathrm{E}}}{-3/16V_{0}} &\simeq& 3.7.
\end{eqnarray}
Therefore, for large $k$, the action will increase almost by four times compared to that of the original Hawking-Moss instanton.

\subsection{\label{sec:gen}General conditions for existence of the solution}

\begin{figure}
\begin{center}
\includegraphics[scale=0.75]{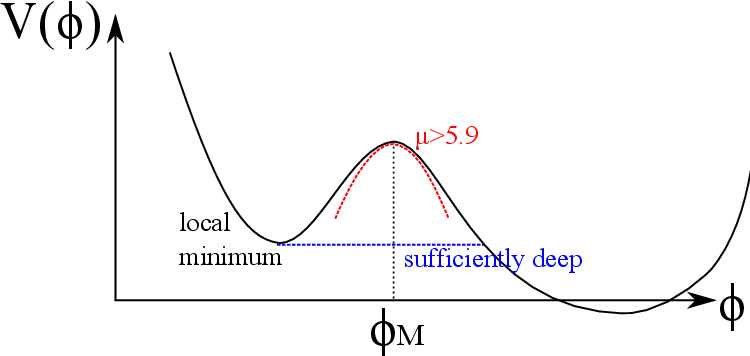}
\caption{\label{fig:field2}Conditions on the potential for the existence of oscillating instantons.}
\end{center}
\end{figure}

Now we are in a position to illustrate the general conditions for the existence of the solutions. Up to now, we considered a double-well potential, especially for the numerical study. However, the analytic discussion of Section~\ref{sec:pro} does not crucially depend on the double-well potential. Therefore, we can reasonably generalize our case to various other potentials (Figure~\ref{fig:field2}).
\begin{enumerate}
\item The potential has a local maximum with $\mu \gtrsim 5.9$.
\item The potential has a local minimum. (Let us say this local minimum as the left side.)
\item The right side behaves as a sufficiently deeper well compared to the vacuum energy of the left side.
\end{enumerate}

The first condition allows for the existence of maximum number of oscillations greater than one; if $\mu$ is greater than $5.9$, then the field begins to oscillate around the local maximum. To satisfy the other two conditions $\dot{\rho}=0$ and $\dot{\phi}=0$ at the same time, we have to tune the initial condition in order to hold the field for a sufficiently long time; the second condition (existence of the local minimum) works with this. However, if the right vacuum is not sufficiently deep, then a trouble might happen: the field begins to roll near the local minimum, and due to sufficient energy, it rolls over the local maximum (i.e., it may not oscillate around the local maximum). The third condition (sufficiently deeper right vacuum) thus makes sure that the field `oscillates' around the local maximum.

These three conditions are quite general and typical for various models. Therefore, these will be extremely useful for various situations.

\section{\label{sec:app}Applications: General vacuum decay problem}

As a simple application of the oscillating instantons, we will consider a vacuum decay problem. There are few previous known examples namely the Coleman-DeLuccia instantons and the Hawking-Moss instantons. In addition, we found that there are yet another kind, the so-called oscillating instantons as homogeneous channels. We will compare these instantons and discuss when one of them dominates.

Let us define the following:
\begin{itemize}
\item $V_{\mathrm{f}}$: Vacuum energy of the false vacuum
\item $V_{\mathrm{M}}$: Vacuum energy of the local maximum
\item $\Delta \phi$: The field distance scale of the thin-wall
\item $T$: The tension of the thin-wall,
\begin{eqnarray}
T \simeq \int_{\phi_{\mathrm{f}}}^{\phi_{\mathrm{f}} + \Delta \phi} d\phi \sqrt{2 V(\phi)},
\end{eqnarray}
and hence approximately given by $\sim \sqrt{2 V_{\mathrm{M}}} \Delta \phi / 2$.
\end{itemize}
Assume that, for a Coleman-DeLuccia type of bubble, the thin-wall approximation is valid with the condition $V_{\mathrm{f}} \ll V_{\mathrm{M}}$. Here, the approximation $T \sim \sqrt{2 V_{\mathrm{M}}} \Delta \phi / 2$ is due to the triangle approximation: $\int_{a}^{b} f(x) dx \sim (b-a)/2 \times \max f$. Of course, it depends on the details of the potential, and hence we need to be cautious for real applications. However, in general, we may intuitively use this approximated relation.

\begin{figure}
\begin{center}
\includegraphics[scale=0.75]{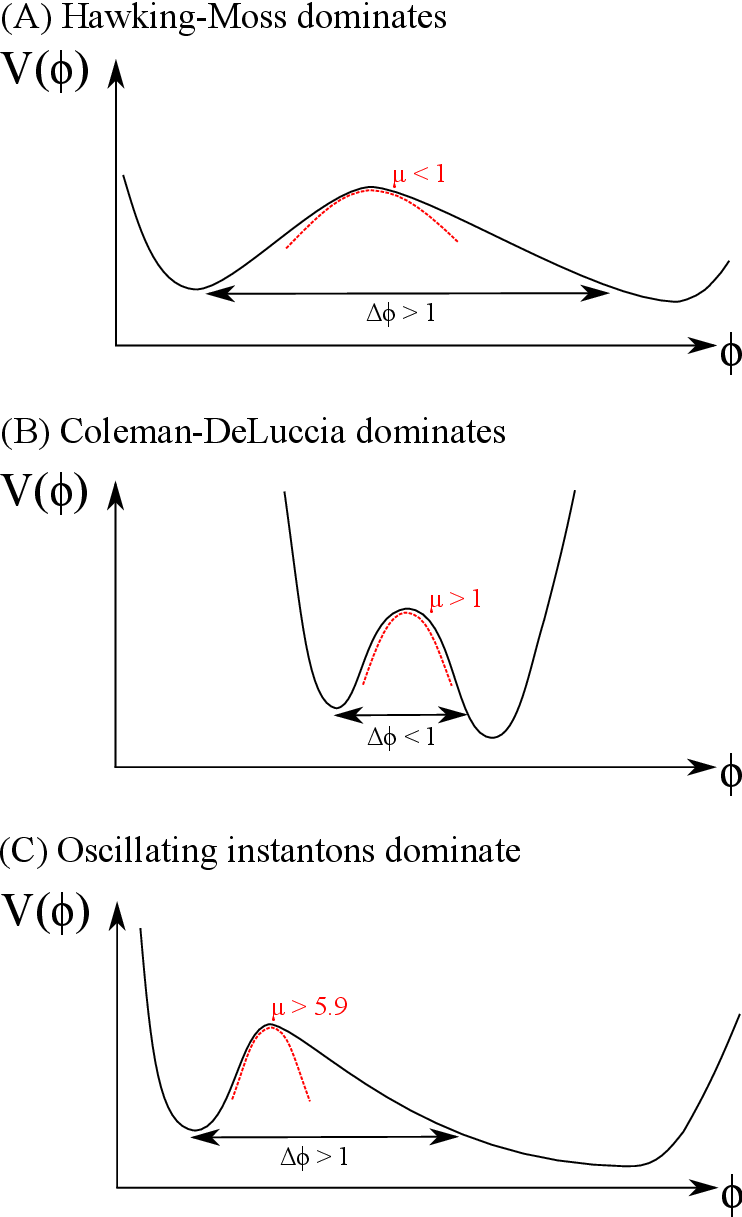}
\caption{\label{fig:interp2}(A) is the Hawking-Moss instanton dominated case ($\mu<1$ and $\Delta \phi >1$). (B) is the Coleman-DeLuccia instanton dominated case ($\mu>1$ and $\Delta \phi <1$). (C) is the oscillating instanton dominated case ($\mu>5.9$ and $\Delta \phi >1$).}
\end{center}
\end{figure}

\begin{figure}
\begin{center}
\includegraphics[scale=0.75]{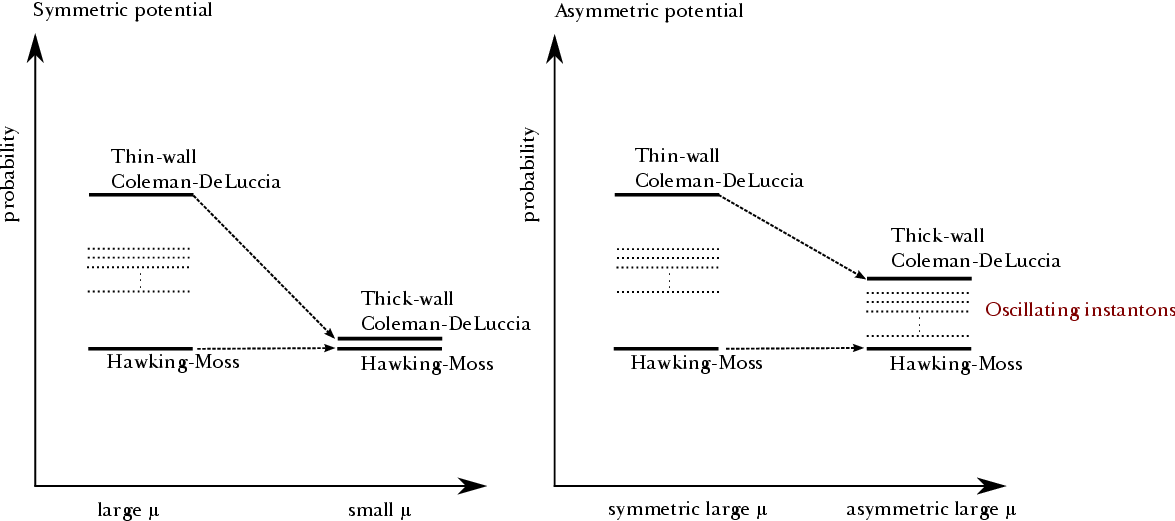}
\caption{\label{fig:conceptual}Conceptual picture of probabilities. Left: When we decrease $\mu$ around the local maximum with symmetry, then the thin-wall Coleman-DeLuccia solution approaches the thick-wall Coleman-DeLuccia solution and this approaches the Hawking-Moss solution. Right: When we change the symmetry with a constant large $\mu$, then the thin-wall Coleman-DeLuccia solution approaches a think-wall Coleman-DeLuccia solution and oscillating instantons do not disappear.}
\end{center}
\end{figure}

The probability for the Coleman-DeLuccia type bubble is as follows \cite{Coleman:1980aw,Kachru:2003aw}:
\begin{eqnarray}\label{eq:CDL}
\log P_{\mathrm{CDL}} \simeq -\frac{3}{8 V_{\mathrm{f}}} \left( 1 + \frac{V_{\mathrm{f}}}{6\pi T^{2}} \right)^{-2}.
\end{eqnarray}
When the gravity is significant enough, i.e., $V_{\mathrm{f}} \ll T^{2}$, we can further approximate this by
\begin{eqnarray}
\log P_{\mathrm{CDL}} \simeq -\frac{3}{8 V_{\mathrm{f}}} \left( 1 - \frac{V_{\mathrm{f}}}{3\pi T^{2}} \right).
\end{eqnarray}
On the other hand, the probability for the Hawking-Moss type instantons is given by,
\begin{eqnarray}
\log P_{\mathrm{HM}} \simeq -\frac{3}{8 V_{\mathrm{f}}} + \frac{3}{8 V_{\mathrm{M}}}.
\end{eqnarray}
Therefore, if we compare these two probabilities, we obtain
\begin{eqnarray}
\log \frac{P_{\mathrm{CDL}}}{P_{\mathrm{HM}}} \simeq \frac{1}{8\pi T^{2}} - \frac{3}{8 V_{\mathrm{M}}}.
\end{eqnarray}
To summarize, we find two limits:
\begin{itemize}
\item If $3 \pi T^{2} > V_{\mathrm{M}}$, then Hawking-Moss type instantons (including the fuzzy instantons) are dominant ((A) in Figure~\ref{fig:interp2}).
\item If $3 \pi T^{2} < V_{\mathrm{M}}$, then Coleman-DeLuccia type instantons are dominant ((B) in Figure~\ref{fig:interp2}).
\end{itemize}
One cautious remark is that in the $3 \pi T^{2} > V_{\mathrm{M}}$ limit, the Coleman-DeLuccia type solutions are no more thin-wall (and hence, Equation~(\ref{eq:CDL}) is no more true). Such type of solutions approaches the thick-wall solutions and the probability eventually approaches the Hawking-Moss instantons (Left in Figure~\ref{fig:conceptual}).

If we include the oscillating instanton solutions, first we notice that the oscillating instanton is always dominant than the original Hawking-Moss instantons:
\begin{eqnarray}
\log P_{\mathrm{osc}} \simeq -\frac{3}{8 V_{\mathrm{f}}} + \frac{3}{8 V_{\mathrm{M}}} \alpha.
\end{eqnarray}
where $\alpha \simeq \mathcal{O}(1) \sim 3.7$ is a numerical factor. Therefore,
\begin{itemize}
\item If $3\pi \alpha T^{2} > V_{\mathrm{M}}$, then oscillating instantons are comparable to Coleman-DeLuccia instantons ((C) in Figure~\ref{fig:interp2}).
\item If $3\pi \alpha T^{2} < V_{\mathrm{M}}$, then Coleman-DeLuccia type instantons are dominant.
\end{itemize}
Again, in the $3\pi \alpha T^{2} > V_{\mathrm{M}}$ limit, the thin-wall formula is no more true and we need cautious comparison. One interesting notice is that in the $\mu>5.9$ and $3\pi \alpha T^{2} > V_{\mathrm{M}}$ limit requires an asymmetric potential.

Let us continue the quantitative discussion with a special example. We use a potential that is similar to the KKLT type \cite{Kachru:2003aw}, although we are only interested in a numerical example:
\begin{eqnarray}\label{eq:pot}
V(\phi) = C_{1} \left( -\frac{C_{2}}{\phi^{2}} e^{-a \phi}+\frac{C_{3}\phi + C_{4}}{\phi^{2}} e^{-2a \phi} +\frac{C_{5}}{\phi^{3}} \right),
\end{eqnarray}
where we choose $C_{1}=6 \times 10^{-5}$, $C_{2}=1000$, $C_{3}=C_{4}=1$, $C_{5}=12.3$, and $a=30$. This potential has a local minimum at $\phi_{\mathrm{m}} \simeq 0.034$ and has a local maximum at $\phi_{\mathrm{M}} \simeq 0.063$. For the large field limit, there is a run-away direction. Around $\phi_{\mathrm{M}}$, we can calculate $\mu \simeq 36.6$. Therefore, this potential can be a good toy model for (C) in Figure~\ref{fig:interp2}.

\begin{figure}
\begin{center}
\includegraphics[scale=0.65]{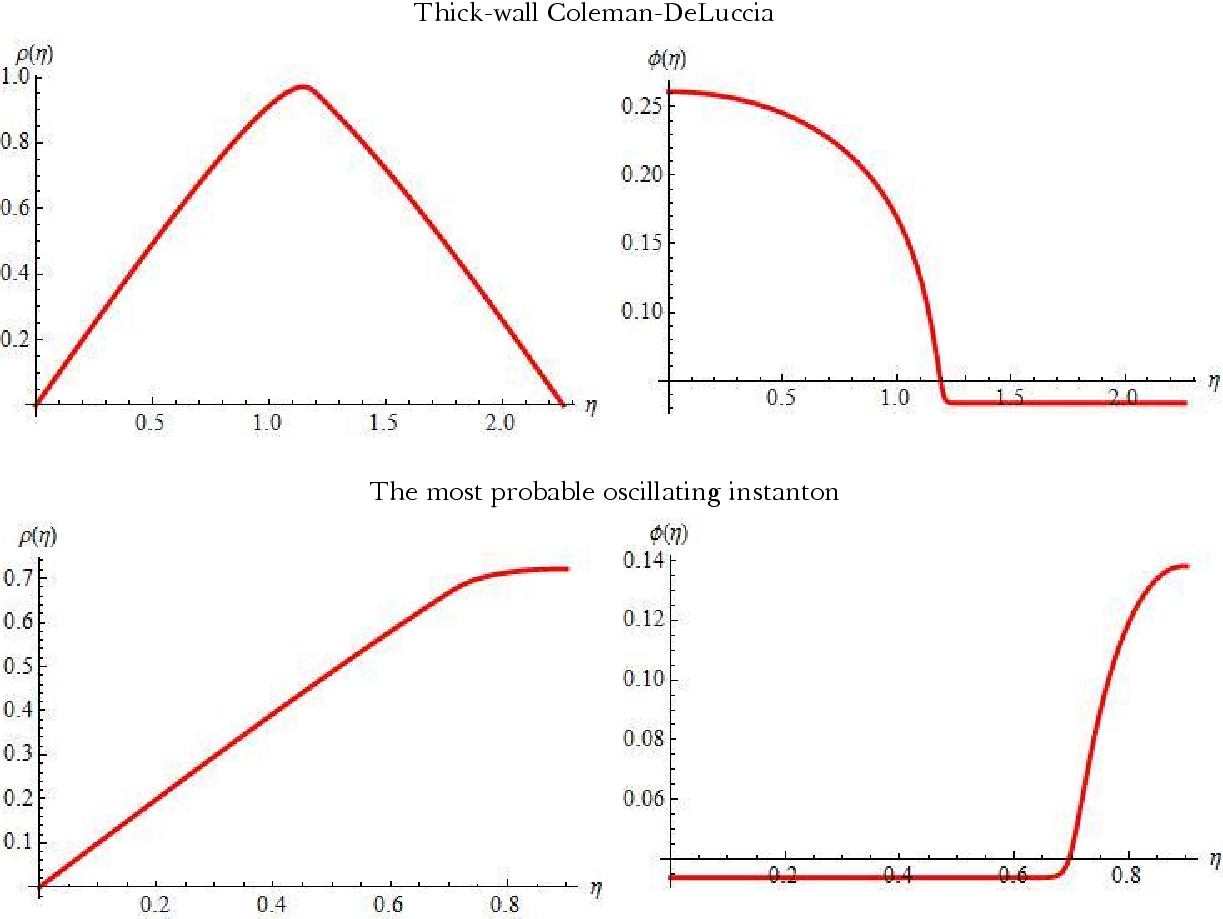}
\caption{\label{fig:results}Numerical solutions of instantons with potential Equation~(\ref{eq:pot}). Upper: Thick-wall Coleman-DeLuccia type instanton. Lower: The most probable oscillating solution $i=0$.}
\end{center}
\end{figure}

One can calculate probability of three instantons (Figure~\ref{fig:results}, we assume that the initial condition is $\phi_{\mathrm{m}}$)\footnote{Note that, in this model, before subtracting the background action ($\simeq - 5.60$), the actions of the solutions are as follows: for CDL, $\simeq - 1.89$, for HM, $\simeq - 0.56$, and for oscillating case, $\simeq - 1.20$. The action ratio between the oscillating case and the HM case is approximately $\sim 2$. If we regard that we used a different potential in Equation (59) rather than Equation (25), this is a consistent result in comparison with Equation (52).}:
\begin{itemize}
\item Coleman-DeLuccia type solution: $\log P \simeq - 3.71$.
\item Hawking-Moss solution: $\log P \simeq - 5.04$.
\item The most probable oscillating solution: $\log P \simeq - 4.40$.
\end{itemize}
Note that if we consider all the oscillating instanton channels, we can approximate that
\begin{eqnarray}
\sum_{i=0}^{N} e^{-B_{i}} \gtrsim N e^{-B_{\mathrm{HM}}} \simeq e^{-B_{\mathrm{HM}} + \log N},
\end{eqnarray}
where $B_{i}$ is the exponent from the Euclidean instanton calculations with $i$-th oscillation, $B_{\mathrm{HM}}$ is the exponent for the Hawking-Moss instanton, and $N$ is the possible maximum oscillation number. In this example, $N \sim 0.17 \mu \sim 6$ and $\log 6 \simeq 1.8$. Therefore, if one adds all the oscillating channels, then they can be more probable than the Coleman-DeLuccia channel.

Figure~\ref{fig:conceptual} summarizes conceptual picture of probabilities. When we decrease $\mu$ around the local maximum and the potential maintain the symmetry, then the thin-wall Coleman-DeLuccia solution approaches the thick-wall Coleman-DeLuccia solution and this approaches the Hawking-Moss solution. In this case, many oscillating instantons (dotted lines) disappear as $\mu$ decreases. On the other hand, when we change the symmetry with a constant large $\mu$, then the thin-wall Coleman-DeLuccia solution approaches a think-wall Coleman-DeLuccia solution, although it does not completely approach the Hawking-Moss instanton. Rather, between two solutions, there remain various oscillating instanton channels. Here, oscillating instantons do not disappear and the probabilities become comparable to the Coleman-DeLuccia channel.

Of course, it does depend on the details of the shape of the potential (or, details of the model) and the choice of parameters. Therefore, it is not easy to make general statements regarding the flux compactification scenarios or string inspired potentials. However, one significant result is that \textit{oscillating instantons are not negligible in general}. We should be more careful about our conclusion whether an oscillating instanton channel is dominant or not, while it is quite common that they have been neglected in many contexts.

One important notice is that there will be a competition between homogeneous oscillating solutions and inhomogeneous oscillating bounce solutions \cite{Hackworth:2004xb}. Although these probabilities are lower than the Coleman-DeLuccia channel, if one adds all the inhomogeneous oscillating bounce channels, then these can be comparable with all the homogeneous oscillating channels. This opens the possibility that the real tunneling process can be more complicated and richer than we traditionally expected.

\section{Discussion}

In this paper, we discuss oscillating instantons as homogeneous tunneling channels.

We assumed Einstein gravity with a minimally coupled scalar field with an $O(4)$ symmetric metric ansatz. We define the oscillating instantons as follows: there exists $\eta_{\mathrm{max}}$ such that $\dot{\rho}(\eta_{\mathrm{max}})=\dot{\phi}(\eta_{\mathrm{max}})=0$.
From this definition, we had two possible analytic continuations to the Lorentzian signature: inhomogeneous way with $\chi \rightarrow \pi/2 + it$ and homogeneous way with $\eta \rightarrow \eta_{\mathrm{max}} + i t$. In our work, we especially focused on the latter case: the so-called oscillating instantons as homogeneous tunneling channels.

We studied the existence of the oscillating instanton solutions using an analytic and numerical techniques. The study of a double-well potential reveals that the existence of an oscillating insanton requires three conditions on the potential. (1) For a local maximum, $\mu > 5.9$ is required. Then there are maximum $N \simeq 0.17 \mu$ numbers of oscillations. (2) For a local minimum, as the initial condition approaches it, the field spends more and more (Euclidean) time around the local minimum of the potential. Therefore, if (3) there exists a sufficiently deep potential barrier on the other side of the potential, then this will make sure that there exists $N$ numbers of oscillating instantons of different kinds by tuning the initial conditions.

We also computed the probability for the oscillating instantons. As the number of oscillation decreases, the probability increases, and eventually has a higher probability than that of the Hawking-Moss instanton. Therefore, if there is an oscillating instanton, in general it is dominant compared to the Hawking-Moss instanton.

We also compared the oscillating instanton with the Coleman-DeLuccia channel. If the potential is sufficiently smooth, then the Hawking-Moss type instantons dominate, while the Coleman-DeLuccia instanton dominats if the potential is steeper. However, if we include the oscillating instanton solutions, we have to consider one more case with $\mu > 5.9$ and $\Delta \phi > 1$, and then the oscillating instantons can be dominant over both the Hawking-Moss and Coleman-DeLuccia instantons. It may be not difficult construct such an example from some toy models of the flux compactification.

For other special models, we should be very careful to apply our calculations. However, what can be definitely said is that \textit{the possibility for an oscillating instanton cannot be simply ruled out}. We should be careful whether to consider an oscillating instanton or not in various special contexts, although it has been neglected in many existing literature.

Finally, we shortly comment on the negative mode issue. Such an oscillating solution can have two or more numbers of negative modes \cite{Battarra:2012vu}. This implies that such oscillating solutions may not be the best way of tunneling \cite{Coleman:1987rm}, in which the author argued that the Euclidean solution with only one
negative mode is related to the tunneling process in the flat Minkowski spacetime. However, there is no rigorous proof on extension of Coleman's argument to the curved space claiming the physical irrelevance of the solutions with additional negative modes. There appears diverse situations on the negative modes when the gravity is taken into account \cite{nemo1, nemo2, nemo4, nemo42, nemo5, nemo52}.
In fact, probabilities of the oscillating solutions are sub-dominant than that of the Coleman-DeLuccia instantons. However, it does not imply that there is nothing to do with oscillating instantons, since they contribute to the Euclidean path integral, although they may not be the peak (steepest-descent) of the wave function. If there are a lot of oscillating instanton channels, then they will contribute some physical effects, and one easy way of interpretation is that these instantons represent thermal excitations \cite{Brown:2007sd}. Therefore, although it is still controversial, it is fair to say that it is worthwhile to study further. We postpone to discuss this issue in future works.

\section*{Acknowledgment}

The authors would like to thank Manu Paranjape and Richard MacKenzie for the hospitality they received during the visit at Universit\'{e} de Montr\'{e}al. The authors also would like to thank for the help of Erick Weinberg, George Lavrelashvili, Dong-il Hwang, Raju and Chaitali Roychowdhury. DY, BHL and WL were supported by the National Research Foundation of Korea grant funded by the Ministry of Education, Science and Technology through the Center for Quantum Spacetime(CQUeST) of Sogang University(2005-0049409). WL was supported by Basic Science Research Program through the National Research Foundation of Korea funded by the Ministry of Education, Science and Technology(2012R1A1A2043908). DY is supported by the JSPS Grant-in-Aid for Scientific Research (A) No.~21244033. We appreciate APCTP for its hospitality during completion of this work.

\end{document}